# PRIVIFY: Designing Tangible Interfaces for Configuring IoT Privacy Preferences

**Bayan Al Muhander**
School of Computer Science and Informatics
Cardiff University, UK
almuhanderb@cardiff.ac.uk

**Omer Rana**
School of Computer Science and Informatics
Cardiff University, UK
RanaOF@cardiff.ac.uk

**Charith Perera**
School of Computer Science and Informatics
Cardiff University, UK
pererac@cardiff.ac.uk

## Abstract

The Internet of Things (IoT) devices, such as smart speakers can collect sensitive user data, necessitating the need for users to manage their privacy preferences. However, configuring these preferences presents users with multiple challenges. Existing privacy controls often lack transparency, are hard to understand, and do not provide meaningful choices. On top of that, users struggle to locate privacy settings due to multiple menus or confusing labeling, which discourages them from using these controls. We introduce *PriviFy* (Privacy Simplify-er), a novel and user-friendly tangible interface that can simplify the configuration of smart devices' privacy settings. *PriviFy* is designed to propose an enhancement to existing hardware by integrating additional features that improve privacy management. We envision that positive feedback and user experiences from our study will inspire consumer product developers and smart device manufacturers to incorporate the useful design elements we have identified. Using fidelity prototyping, we iteratively designed *PriviFy* prototype with 20 participants to include interactive features such as knobs, buttons, lights, and notifications that allow users to configure their data privacy preferences and receive confirmation of their choices. We further evaluated *PriviFy* high-fidelity prototype with 20 more participants. Our results show that *PriviFy* helps simplify the complexity of privacy preferences configuration with a significant usability score at p < .05 (P = 0.000000017, t = -8.8639). *PriviFy* successfully met users' privacy needs and enabled them to regain control over their data. We conclude by recommending the importance of designing specific privacy configuration options ([Demo Video](#)).

*Keywords* Data Privacy; Privacy Awareness; Privacy Preferences; Privacy Configurations; Privacy Choice; Physical visualization; Ambient Display; Sensors; Internet of Things (IoT); Usable Privacy; Design Space

## 1 Introduction

The connection between using consumers' data and IoT technology is imperative, as most IoT devices depend on this data to operate [1]. The extensive data collection capability of IoT devices, however, has prompted notable privacy concerns among users [2]. As a countermeasure, some IoT companies have introduced privacy choice mechanisms, such as controlling data retention period and targeted advertising opt-outs, asserting that users possess authority over their data [3]. In contrast to these assertions, the majority of the available privacy controls demonstrate inadequacies in meeting users' privacy needs [4].

Privacy research indicates that even with privacy controls in place, the privacy choice mechanism may not effectively protect consumer privacy [5, 6]. This is partly because of the challenges associated with the poor usability of privacy choice mechanisms [3]. In many cases, privacy controls lack transparency and fail to offer meaningful choices [7]. This is further complicated by vague explanations of privacy-related concepts, which often hinder users from locating and utilizing them effectively [8]. The design of a privacy choice interface plays a crucial role in shaping users' privacy outcomes, particularly in the context of IoT [9, 3]. Unlike typical user interfaces, privacy choice interfaces in the IoT demand special usability considerations [9, 3]. Generally, most IoT users are not tech-savvy individuals and might only consider configuring their privacy settings if the choice interface is engaging to use [10, 11].

Numerous privacy research studies have delved into developing usable and transparent mechanisms for privacy choices, aiming to empower users to exercise their consent and choice options [12, 13]. Despite these efforts, many existing methods face challenges in terms of adaptation, user engagement, and practicality. For instance, setting up privacy preferences in some approaches requires users to navigate through multiple menus and options, which can be complex and requires technical expertise [14]. Figure 1 illustrates composite representations of the privacy settings interfaces across various IoT applications, highlighting the diverse array of menus and options that users encounter while configuring their privacy preferences. Other approaches also require users to read explanations, which most users tend to ignore [15]. In addition, most privacy choice interfaces are often software-based, making them more complex and less intuitive [16]. These factors collectively hinder their effectiveness and user acceptance.

In response to these challenges, *tangible privacy* emerges as a concept proposed to enhance individual privacy within the IoT landscape [17, 11]. Tangible privacy enable users to directly and tangibly interact with their privacy settings, presenting a promising avenue for improvement [11]. Studies have demonstrated that tangible privacy enhances users' interactivity and engagement, providing a more user-friendly approach to managing privacy preferences [18, 19]. In Table 1, we provide a comparison of multiple tangible privacy studies by examining various aspects, including configuration methods, feedback mechanisms, privacy settings configuration, software interface independence, configurable IoT devices, and integration into existing systems. By evaluating these factors, this study builds on prior research on usability aspects to design an interface that empowers users to regain control over their privacy and make informed choices [3, 20].

We developed *PriviFy*, a tangible interface facilitating direct interaction with privacy preferences and enhancing user understanding of available options. *PriviFy* prototype presents the possibility of leveraging existing hardware, such as routers or Amazon Echo, to serve as facilitators of enhanced privacy management. Our prototype implementation employs various physical components to validate our approach's feasibility and usability. We used knobs and buttons with concise, descriptive text to facilitate user interaction, enabling users to easily correlate them with specific privacy choices. In order to give users meaningful feedback on their configuration, we integrated lights and short confirmation messages on the display. This approach aims to streamline the process of configuring privacy preferences for IoT, offering users an intuitive interface for managing their privacy settings. The proposed solution seeks to enhance user control over data privacy while eliminating the need to navigate through multiple applications or interfaces. The physical controls will provide tactile feedback, enhancing users' experience and making it easier for users to adjust privacy settings according to their preferences. The current study makes the following research contributions:

- We performed multiple rounds of focus group studies to gain insights into individuals' understanding of privacy and to understand the complexities inherent in the current privacy configuration options. Throughout each round, we iteratively refined and informed the design of the prototype based on the insights gathered.

- We present the design and implementation of *PriviFy*, an interactive tangible interface that simplifies managing IoT privacy preferences for individuals, supporting them to gain power over their privacy and providing feedback about their choices. *PriviFy* interactively assists IoT users in managing and understanding their privacy preferences.

- We report the findings of our study consisting of a set of semi-structured interviews conducted with twenty participants using *PriviFy*. These results demonstrate the potential of a tangible interface to encourage users to manage their privacy.

The remainder of this paper is structured as follows: in Section 2, we describe related work about IoT privacy concerns, privacy choices usability, and tangible privacy. Section 3 includes a description of the proposed prototype design methodology. Section 4 presents the different methodology phases to implement *PriviFy's* prototype. We describe the evaluation in Section 5. Section 6 presents the research findings, and Section 7 discusses the research challenges and future opportunities, with concluding comments in Section 8.

## 1.1 Motivation

Current research on privacy choices predominantly focuses on web and mobile interfaces, with less attention given to the IoT [28, 29]. Implementing meaningful privacy choices in the IoT presents several challenges [30], with a significant obstacle stemming from the practical constraints of limited user interfaces [31]. Configuring privacy settings for IoT devices often proves to be a cumbersome and unintuitive process, typically requiring users to navigate complex menus and settings within the associated mobile applications. In light of these challenges, and inspired by the tangible user-friendly interfaces of devices such as the Stream Deck [32], as well as the intuitive physical controls of ATMs, we adopted tangible interfaces due to their inherent advantages in enhancing user understanding and engagement. Figure 2 depicts the inspiration behind the proposed prototype for simplifying privacy configuration of IoT devices.

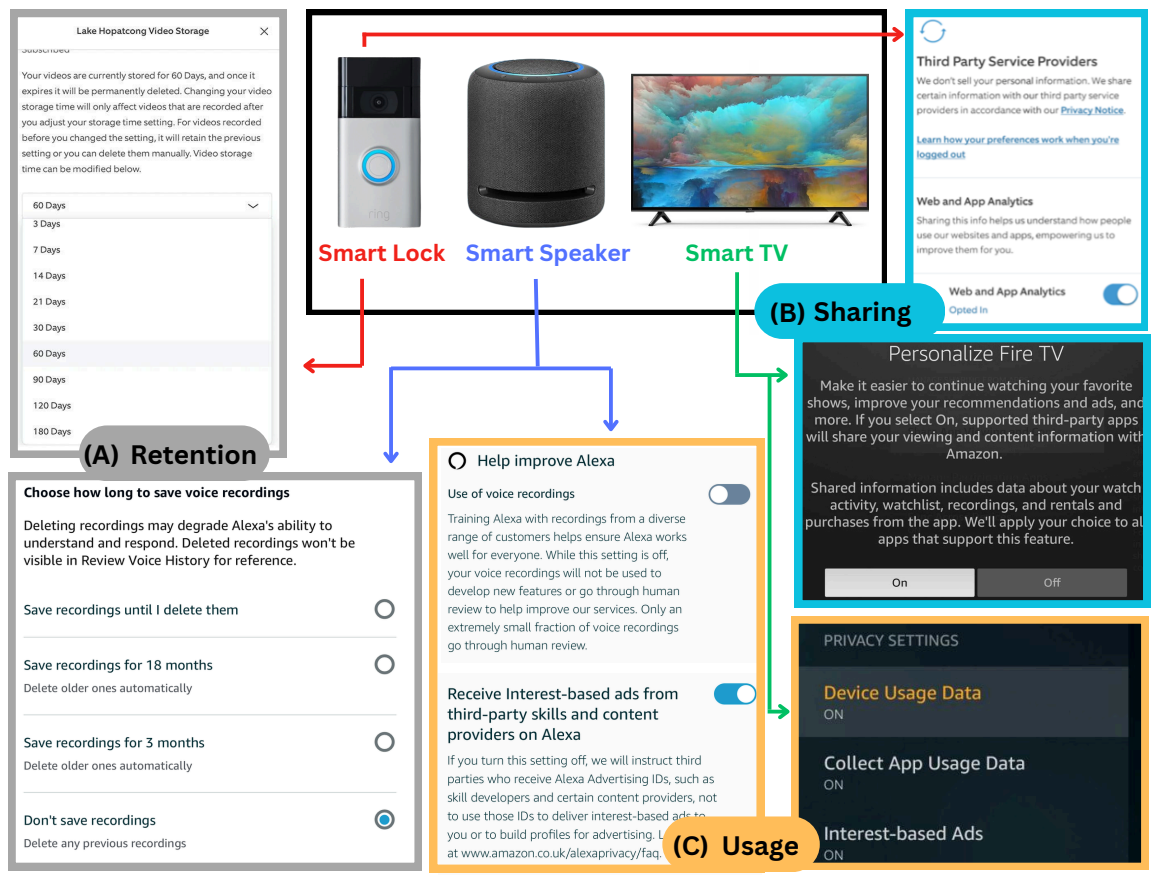

Figure 1: A composite representation of the varying menus and options available on three smart devices (Smart Lock [21] , Smart Speaker [22], and Smart TV [23]) depicting each device's privacy settings interface. (A) shows the privacy settings for retaining data, (B) shows the privacy settings for sharing data, and (C) shows the privacy settings for using data.

Our prototype aims to address the complexities of managing privacy preferences in the IoT context by providing a simple and engaging solution. By integrating privacy management functionalities into existing hardware, complemented by physical controls such as knobs and buttons, users can intuitively adjust privacy settings without the need for complex software interfaces [26]. This approach aims to empower users to have control over their privacy. It also lets users interact physically with their privacy configuration, providing a more intuitive and user-friendly interface. Our solution goal is to simplify the configuration process into abstract capabilities. Rather than offering an exhaustive array of settings, we prioritize usability by abstracting fine-grained configurations into manageable options. As confirmed by the results of our evaluation, this method ensures that users can easily navigate and understand their privacy settings without overwhelming complexity. By acknowledging the need for simplicity, our solution balances user control and ease of use, ultimately enhancing privacy management for IoT devices.

Table 1: Comparison of Tangible Privacy Prototypes to Highlight the Unique Features of *PriviFy*.

| Privacy Device | Method of Configuration | Feedback Mechanism | Privacy Configuration | Tangible Exclusivity | Configurable IoT Devices | Integration |
|---|---|---|---|---|---|---|
| PriKey [16] | Sliders & Button | Light | Limited | ✗ | Multiple Devices | ✗ |
| SNOP [24] | Moving Disk | Haptic | Limited | ✓ | Single Device | ✓ |
| Dashboards [24] | N/A | Colored Pins | None | ✓ | Multiple Devices | ✗ |
| Ambient Lighting [25] | N/A | Light | None | ✓ | Single Device | ✗ |
| Privacy Speaker [25] | N/A | Audible | None | ✓ | Single device | ✗ |
| *Ahmed et al.* [26] | Sliding Button | Physical | Limited | ✓ | Single Device | ✓ |
| *Ahmed et al.* [26] | Sliding Button | Light | Limited | ✓ | Single Device | ✓ |
| Privacy Band [27] | Haptic Touch | Haptic | Limited | ✗ | Single Device | ✗ |
| *PriviFy* | Knobs & Buttons | Light, Haptic, & Visual | Moderate | ✓ | Multiple Devices | ✓ |

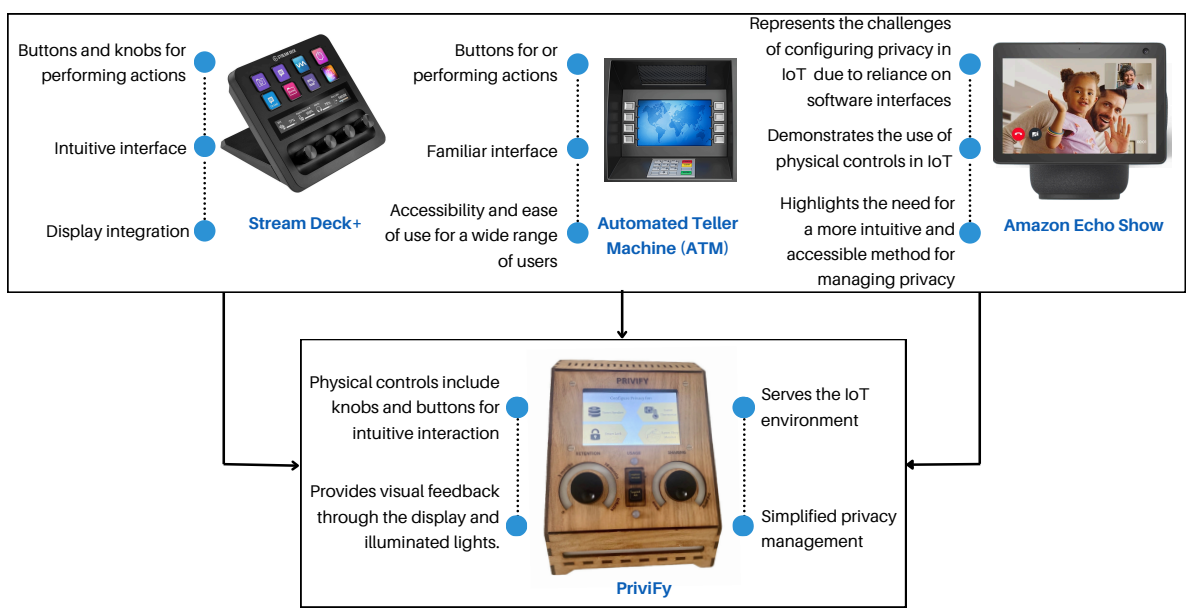

Figure 2: PriviFy, inspired by the user-friendly interfaces of Stream Deck and ATM, as well as the challenges presented by the Amazon Echo in managing privacy settings. By leveraging physical controls such as knobs and buttons, PriviFy offers a seamless and accessible solution for managing privacy settings in IoT environment.

## 2 Related Work

### 2.1 IoT Privacy Concerns

IoT frequently transfers user data to a growing number of service providers, raising significant privacy concerns [33, 34]. Numerous studies have highlighted users' concerns about the potential privacy risks associated with collecting, processing, and sharing personal data by IoT devices [33, 35, 36]. In light of these concerns, some IoT solutions provide users with configuration options to control their data [37, 38, 39, 40]. However, the complex structure of IoT device configuration options often makes it challenging for users to understand and control their data [9, 41].

Options for privacy configuration tend to require extensive reading, are not obvious or employ ambiguous terminology, making them difficult to locate or understand [8, 7]. In addition, system architectures based on IoT can often track locations, detect the environment, and collect and store data that may be considered private [42]. This capability of IoT systems, coupled with the difficulty of controlling data, raises substantial concerns among users regarding the potential infringement of their privacy. Studies have consistently observed users' uncertainty about who has access to their personal information and for what purpose, as well as their feelings of helplessness regarding controlling their data [25, 43, 44]. As users become increasingly aware of these privacy challenges associated with IoT, there is a growing demand for transparent data practices, user-friendly control mechanisms, and efficient privacy management solutions to mitigate these concerns and establish a sense of trust in IoT technologies [9].

### 2.2 Privacy Choices Usability

A growing body of research has investigated various aspects of creating privacy choice mechanisms to tackle usability issues [9, 45, 46, 47]. One of the earliest examples is the "Platform for Privacy Preferences Project" (P3P) [48]. P3P emerged to assist users in managing the release of their data by blocking websites that do not comply with their privacy choices [48]. As a result, users will not need to read lengthy privacy policies [48]. Following P3P, several web, mobile, and IoT tools targeted at improving usability were developed [28]. Notable examples include Privacy Bird [49], which enables users to specify recipients, data categories, purposes, and retention. The Nutrition Label for privacy [50] focuses on recipients, categories, and purposes and has been enhanced in the Visual Interactive Privacy Policy (VIPP) [51] and Privacy Policy Options (PPO) [52] to give users control over data retention and deletion. PeopleFinder [53] was developed as an application to facilitate location sharing and allow users to specify the length of such sharing. The IoT Assistant application (IoTA) [9] empowers users to set their privacy preferences when applicable. The Privacy Badge [54] visually represents the nature of disclosed data, the timing, the recipient, and the purpose of disclosure, where users are given the capability to configure their privacy preferences through this visualization. Additionally, the DigiSwitch [55] medical system employs a digital photo frame to display collected data and provides users with the option to temporarily halt data transmission.

Although there is increasing recognition of usability issues, existing methods for managing privacy preferences continue to face significant usability challenges [5, 46, 56]. For instance, some studies indicated that users might be misled by the terminology and ambiguous expressions when attempting to manage their privacy preferences [57, 58]. Additional research also emphasized the complexity and lack of user-friendliness in existing privacy settings interfaces [59, 60]. Users frequently encounter difficulties navigating through complex menus and understanding the implications of different privacy options [61, 56]. This usability gap not only contributes to privacy challenges but also impedes users from making informed decisions about their data [62]. To address this issue, there is a crucial need for interventions that enhance usability and improve the user-friendliness of privacy choices, encouraging users to actively engage in managing their privacy settings [56]. In the following sections, we outline four essential design considerations that inform our proposed interface for user privacy preferences. We leverage these design recommendations from previous research on the domain of design space [3].

### 2.3 Tangible Privacy

In response to the challenges associated with privacy interfaces' usability, literature has proposed tangible privacy interfaces as a promising solution [19, 18, 63]. Tangible interfaces are defined as physical, touchable interfaces that allow users to easily manage their privacy settings through tangible interaction, affording them a more intuitive and interactive experience [11, 64, 19, 17]. The use of tangible interfaces has been exploited across a variety of fields, including encouraging playfulness and curiosity [63], promoting communication and socialisation [65, 66], and inspiring the exploration of new topics through learning [63, 64]. In privacy research, tangible interfaces have been used to enhance privacy usability and foster user interaction with their data [16, 24, 18].

Studies suggest that tangible interfaces can bridge the gap between users and complex privacy settings by providing a more intuitive, interactive, and user-friendly representation of privacy choices [11, 24]. This physical engagement can enhance users' understanding of their configuration options and motivate active management of their privacy settings [19]. Hence, by integrating tangible privacy interfaces, we aim to create a more usable approach to IoT privacy management, fostering a sense of control and trust among users.

## 3 Design Consideration and Process

Our methodology for creating *PriviFy* prototype consists of four key stages. In the first stage, a thorough literature search is conducted to inform the design, followed by an explicit definition of the problem. The second stage involves careful consideration of design principles and formulating a structured design process. In both the first and second stages, we had an extensive discussion with privacy professionals and product designers to inform the design. The third stage consists of iteratively creating low-, medium-, and high-fidelity prototypes, with each stage refining the design based on insights gained. Finally, the final stage involves user testing and evaluation to assess the high-fidelity prototype's effectiveness, usability, and user satisfaction, thereby validating the proposed design against predefined criteria. This systematic approach ensures a comprehensive and iterative progression from problem identification to prototype development to validation while adhering to established design considerations.

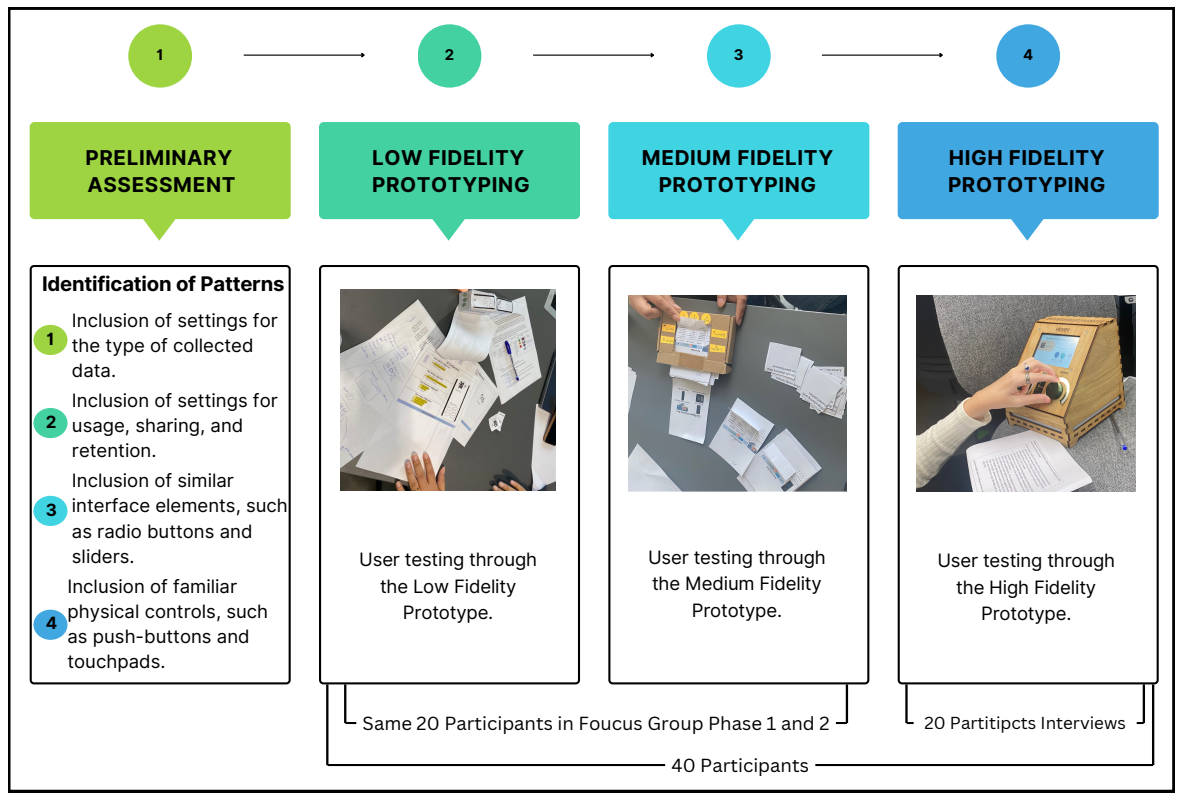

Figure 3: Design process of *PriviFy*, from preliminary assessment to high-fidelity prototyping. (1) Preliminary assessment, involving identification of patterns. (2) Low-fidelity prototype, developed through focus groups with 20 participants, facilitating early-stage exploration of design concepts. (3) Medium-fidelity prototype, refined based on feedback from additional focus groups with the same 20 participants, ensuring iterative improvement. (4) High-fidelity prototype, validated through semi-structured interviews with 20 participants, confirming usability and effectiveness.

## 3.1 Design Considerations

Drawing upon recommendations from prior work on design space, we identify four key design considerations [3]. These considerations guide the design of our privacy configuration interface.

- Usability: The interface should enable users to perform the action they intend to perform with minimal effort.
- Findability: The interface should make it easy for users to locate the specific privacy controls they require, regardless of their level of knowledge or expertise. Ideally, users should quickly regain proficiency in the interface after a period of inactivity.
- User Engagement: The interface should effectively draw users to interact with it and provide a pleasant and satisfying user experience.
- User needs: The interface should provide users with privacy choices that are consistent with their privacy preferences when making privacy decisions.

## 3.2 Design Process

We initiated the design process for *PriviFy* by thoroughly examining IoT devices that offer any type of tangible interaction. We evaluated 100 IoT devices, of which only 30 have physical control capabilities; refer to the supplemental file for the full list [67]. Typically, other IoT devices are managed entirely via a software application. Our selection strategy was inclusive, to capture a representative sample of the IoT landscape so that we could conduct a thorough analysis of privacy settings and control mechanisms. We identified a wide variety of IoT devices in several categories, including smart home appliances, wearable technology, health monitoring devices, and environmental sensors. We selected devices based on their prevalence and popularity in the consumer market, with a focus on devices that allow for tangible interaction. In addition to reviewing the 30 devices, we looked at the mobile applications associated with each device, where applicable, to learn about their privacy settings and configuration control. Throughout our review, we noticed the following patterns:

- Existing privacy policies specify what data is collected and how it can be configured. The tangible interface then should then address data collection configurations.

- Existing privacy policies include settings for data usage, sharing, and retention. Following that, the tangible interface should reflect those settings.

- Existing privacy policies include radio buttons, action buttons, toggles, sliders, drop-down lists, and checkboxes for configuring privacy settings. Therefore, to reduce the complexity of the privacy topic, the tangible interface should incorporate similar forms.

- Existing IoT devices provide physical control in the form of push-buttons, switches, sliders, touch pads, and control Knobs. Subsequently, the tangible interface should incorporate similar forms to gain higher user acceptance.

Following the identification of patterns, we used fidelity prototyping. This process engaged 40 participants to iteratively create an interactive prototype design, described in the next section. Figure 3 depicts the design process of PriviFy, from preliminary assessment to high-fidelity prototyping.

## 4 Methodology

We conducted two focus group studies and twenty semi-structured interviews at the University department. The two focus group studies comprise the same twenty participants, 12 females and 8 males, recruited via social media platforms. Our semi-structured interviews included twenty participants, consisting of nine males and eleven females. These participants were recruited through mailing lists. Before their participation, each participant received an information sheet and completed a consent form. Each focus group session had a duration of one hour, which included a snack break. The duration of the semi-structured interviews with each participant averaged one hour. The participants were free to leave the study at any time. With the participants' permission, the moderator audio-recorded each study session and made occasional pictures to document the study. We conducted both studies after receiving ethics approval.

### 4.1 Low Fidelity Prototyping

Based on the findings in Section 3.2 and insights derived from current tangible interfaces, we created three low-fidelity prototypes for *PriviFy*: (1) A cuboid shape with five sides, (2) A table layout with a touch-pad, (3) A table layout with a card reader. The prototypes are described in Appendix A and depicted in Figure 8 (B). The objective of this focus group was to generate ideas for a potential physical interface and collect feedback on *PriviFy's* initial low-fidelity designs.

We started by providing the participants with an explanation of the existing IoT privacy policies and the types of configurations they offer. We engaged in a discussion regarding the difficulties individuals encounter when attempting to manage their privacy preferences. Next, we presented the participants with various images of tangible interfaces and discussed the suitability of constructing a tangible interface for configuring IoT privacy preferences. Subsequently, we divided the participants into two groups and provided them with a guidelines page, pens, and a prototyping sheet, see Figure 8 (A). They were then instructed to engage in a brainstorming session, drawing upon their ideas to collaboratively create a low-fidelity tangible interface design aimed at empowering individuals to manage their privacy settings. Each group developed different prototypes from the supplied materials, as shown in Figure 8 (C). Lastly, we presented our three low-fidelity prototypes, shown in Figure 8 (B), to gather input from participants.

### 4.2 Medium Fidelity Prototyping

Based on the findings obtained from the low-fidelity prototyping session, we systematically improved the design of *PriviFy* and developed two medium-fidelity prototypes: (1) A design resembling an ATM, and (2) A design resembling a control panel. The prototypes are described in Appendix A. This subsequent study aimed to enable individuals to visually perceive and interact with the tangible interface, and gather their insights on how we can develop the high-fidelity prototype.

In this session, we divided the participants into pairs and provided each pair with four different use cases of IoT devices, refer to Figure 11 (A) in Appendix A. We presented each use case with a brief overview of the IoT device outlining its data processing practices for capturing, using, sharing, storing, and retaining the users' data. Participants were then asked to specify which IoT device they wanted to configure, providing rationales for their choices, (Figure 11 (B)). Subsequently, we introduced the two medium-fidelity prototypes, which we created using stickers, papers, and cardboard boxes. We instructed the participants to use these prototypes to configure their preferences. Figure 9 depicts the two medium-fidelity prototypes. Throughout the process, participants were encouraged to think aloud, explaining the reasoning behind their selected configurations. In addition, we asked the participants to identify any aspects they found favourable or unfavourable about the two prototypes and propose enhancements to improve usability. This enabled a thorough assessment to improve the prototype's usability.

### 4.3 Low and Medium Fidelity Prototyping Results

We transcribed the two sessions' audio and combined them with the occasional photographs, the participants' prototypes, and the participants' feedback on our prototypes. To enhance clarity, we replicated the designs of the participants' low-fidelity prototypes using a computer text editor, as seen in Figure 10 in Appendix A. We obtained comprehensive insights into how individuals perceive and set their privacy preferences from the data. Based on the data, we distilled 30 initial codes that were further categorized into four primary themes: **(i)** Content and Design, **(ii)** Preferences Control, **(iii)** Usability, and **(iv)** Awareness of Choices. Within the next part, we discuss the prevailing themes that arose from our qualitative study. The analysis of the prototyping discussion yielded qualitative insights indicating that participants encountered challenges when engaging with their privacy preferences. All participants agreed that the current depiction of privacy choices fails to meet their needs. Participants indicated that having a dedicated tangible device designed to handle privacy would motivate them to engage with their privacy.

**Content and Design:** In the first session, both groups collaboratively designed a flat-shaped prototype with buttons and touch screens, emphasizing the familiarity of these elements to enhance user engagement in controlling privacy preferences. Participants highlighted that using familiar language (i.e., text and icons) would give the tangible interface a more neutral appearance and make it easier to use. *"We've added the sharing and retention icons because we see them everywhere and understand what they mean, refer to Figure 10 (A)" P2 and P3 commented.* The cuboid prototype, while perceived as enjoyable, was deemed unfamiliar by most, with differing opinions on its potential educational value for children. The card reader feature received criticism, but the table interface, featuring a touchpad and swiping for displaying IoT device data, was well-received. Some participants recommended maintaining consistency in interface size, and all participants agreed on the effectiveness of colour changes as a visual indicator for configuration intent.

In the second session, all participants expressed a desire for a more usable configuration option for smart speakers, six for the smart lock, and only four for the sleep monitor and thermostat. Participants commented that because the smart speaker and smart lock can capture identifiable data, such as audio and video data, they must have more usable configuration choices. Concerns about data usage, retention, and sharing were common, with participants posing questions like, *"How is my audio used? Is it shared with a third party, or is it saved forever? I wish I could modify the duration of data storage."* All participants liked the knob on the control panel prototype, and six suggested reducing the options in the data sharing knob. Most participants preferred the control panel prototype layout over the ATM layout. Eight participants recommended an angled touch screen with a stand for the tangible device, which can facilitate placement on home shelves. While swiping right and left was generally preferred, participants acknowledged that selecting devices from the screen might be more intuitive. In response to user feedback, participants proposed changing the button colours or adding indicator lights to communicate data usage. The delete-all option was well appreciated and deemed quite beneficial by all participants. These findings provide useful insights into user preferences and concerns about smart devices' settings, emphasising the significance of user-friendly interfaces in this context.

**Preferences Control:** During the first session, participants appreciated the concept of controlling their privacy preferences through a tangible interface, where both groups designs resembled a small phone frame that could be placed in an easily accessible location. Emphasising the need for quick access, participants noted the time-consuming nature of configuring diverse IoT devices individually and expressed anticipation that a tangible interface would simplify this process. Aligned with the design goals, participants sought a practical design highlighting the significance of easy and quick findability in influencing user engagement with privacy configuration. Participants also tried simplifying the prototypes by including only the essential configuration options and emphasising the effectiveness of choices, *"We tried to include only the things that raise concerns to avoid overflowing the design but still provide effective options."* P1, P3, and P4 stated that the number of options in the prototypes created by the research team is reasonable and logically structured.

During the second session, participants noted the lack of an interactive privacy configuration interface for the four use cases' devices. One participant (P11), reflecting on their personal experience with a smart speaker, expressed the challenge of navigating the application to configure privacy settings, stating, *"I always want to limit how long my audio is saved, but I find it difficult to log in to the application and find the right configuration, so I ignore it. A tangible device like this would make it much easier."* Participants enjoyed the idea of having physical controls, such as knobs and buttons, noting that the physical interaction with these controls can be intuitive and could encourage device usage. Moreover, participants also emphasised the importance of simplicity in configuration options as several participants commented on the perceived complexity of the ATM prototype, expressing that too many buttons could be overwhelming. Participants suggested options categorisation and limiting the options to a maximum of three to enhance user experience. Participants strongly preferred the physical interface concept, noting how it can substantially simplify the process of locating privacy choices and recognising its potential to reduce user effort.

**Usability:** One of our design goals is to create a usable prototype that simplifies the privacy topic and requires minimal user effort. Throughout the first session, participant feedback affirmed the simplicity, with statements such as *"I liked how simple it is to swipe the screen to configure my privacy preferences; it doesn't require any work,"* P9 noted. *"It's convenient having all my important configurations physically in front of me, all I have to do is click a button!"* commented by P15. Prototypes developed by the participants, see Figure 10, also employed a simple layout with three simple control buttons. In addition, participants discussed the convenience of the tangible interface in sharing control, particularly in the effortlessness of managing preferences and the potential for user delegation of control. For instance, P5 said, *"I share my camera recordings with my mother. Sometimes I want to stop sharing for a while, but managing the preferences is tedious, so I leave it on all the time. This physical interface will make stopping the sharing easy. I can even ask my kids to click a button to stop sharing."*

Throughout the second session, participants consistently reaffirmed the practicality of the tangible interface concept, highlighting its seamless facilitation of privacy preference management. Participants responded positively to having such a physical device in their homes, preferring it over web or mobile interfaces. A major focus of their reviews was its usability, specifically how it blends into the home environment, unlike mobile apps that may go unnoticed in the background, and its solely focus on privacy without introducing unnecessary distractions. The knob rotational control element received positive feedback, with four participants suggesting the addition of lights with each rotation. Additionally, participants recommended limiting the focus to crucial data, particularly identifiable data such as location, visual, and audio. This shared perspective demonstrates the perceived advantages of a dedicated, tangible interface in enhancing user engagement with privacy settings in the context of smart devices.

**Awareness of Choices:** Throughout the study, participants engaged in discussions about their limited understanding of the available IoT privacy preferences choices. They also argued that although the research team's prototypes provide easily accessible and findable preferences, they do not explain how adjusting privacy settings influences data collection and use, particularly what can be learned from the collected data. Moreover, participants stated that the prototypes do not communicate the outcomes of an individual's decisions. *"I believe that the prototype should include what happens if I select specific configurations, such as whether I will have limited access to the service or not,"* P5 said. Participants emphasized that the prototype should be both concise and understandable to facilitate informed decision-making. In response, participants suggested adding a popup window feature that appears when configuring privacy settings, as depicted in Figure 10 (B). This popup provides information about the type of data collected and the potential impact of each privacy choice, aiming to enhance participants' understanding for more informed decision-making.

We modified our medium-fidelity prototypes to respond to participants' feedback on the need for clearer information on privacy choices. We introduced a popup window that appears during privacy settings configuration, aiming to raise awareness about the collected data and the potential consequences associated with each privacy choice. During the second session, participants interacted with our prototypes and noticed the increased information available to them about their choices. For instance, a popup explaining data retention prompted one participant (P3) to say, *"Aha, now I know why the smart speaker needs to store my data for a long time!"* Participants also acknowledged the addition of explanations following each adjustment to privacy settings. Participants suggested including a brief device description after selecting each device, to provide a quick reminder of its capabilities. This iterative process reflects our systematic approach to improving user understanding and supporting informed decision-making regarding privacy settings.

## 4.4 High Fidelity Prototyping

After conducting multiple rounds of user studies following the design process outlined in Section 3.2 and considering the four design considerations detailed in Section 3.1 (namely, **(i)** User Need, **(ii)** User Engagement, **(iii)** Usability, and **(iv)** Findability), we derived several design modifications from our low and medium fidelity prototyping sessions. These modifications aim to enhance the tangible interfaces to ensure their general adoption as a viable means for configuring privacy preferences. The following explains the modifications and how we incorporated them into the high-fidelity prototype. Figure 4 presents *PriviFy* high fidelity prototype. We also present a fictional use case scenario involving an individual named Sara engaging with a smart speaker to show how *PriviFy* works and how individuals can interact with it to configure their privacy preferences.

**User Needs:** Our participants strongly desire a device that fulfils their needs and addresses their concerns. They emphasized the importance of having essential options and the ability to get information about each choice. In our prototype, we addressed this by adding the three most important options as suggested by the participants (retention, usage, and sharing), refer to Figure 4. We also added notification messages to inform users about their choices, refer to Figure 12 (D). In addition, we included descriptions for each choice, explaining what data the device collects and its significance. To give users more control, we allowed them to tap on the specified data type and then control it using the physical knobs or buttons.

**User Engagement:** The results from our low- and medium-fidelity prototyping sessions suggest that for the tangible interface to attract users and deliver a pleasant user experience, it should integrate seamlessly into its intended environment. Consequently, we incorporated wood into our high-fidelity prototype, as we anticipate testing it with smart home users. Given wood's common use in home furniture, we aimed to enhance user acceptance [68]. We included the most liked physical controls, i.e., knobs and buttons, carefully selecting appropriate screen size and control dimensions and colors to maintain consistency across the interface elements. We opted for an angled design, facilitating easy placement on a home shelf for convenient participant engagement.

**Findability:** One of our purposes in developing a tangible interface is to offer users an intuitive method for adjusting their privacy settings. We organized the IoT devices on the touch screen in a clear and large format, making it easy for users to locate the specific device they wish to configure (refer to Figure 12 (A)). Responding to participants' feedback, we incorporated touch functionality to select the device on the home screen instead of swiping, aiming to simplify the process and make it more intuitive. Once users choose a device for configuration, they see a concise yet informative detail about the device, along with a clearly organized set of circled data types for quick identification and selection (see Figure 12 (B)). Following selecting the data to configure, users can read a message directing them to use the physical controls for the configuration process (see Figure 12 (C)). We anticipate that this interface design will facilitate users in easily locating specific privacy controls, irrespective of their level of knowledge or expertise, and allow for quick proficiency regain even after a period of inactivity.

**Usability:** Through our iterative design process, participants consistently preferred a simple interface with limited options. This aligns with prior research advocating for the inclusion of only fundamental and logically structured configuration choices [69]. Options that were emphasized were related to data types, retention, usage, and sharing. Our high-fidelity prototype addresses this by including three categorizations—retention, usage, and sharing— logically organized below the screen. We allowed users to select data types directly on the screen to enhance user interaction and reduce complexity. Additionally, we added short descriptions around the knobs and buttons with lights for immediate feedback, aiming to simplify the user experience and minimize effort.

### 4.5 *PriviFy* Functionality (Use Case Scenario)

We implemented this use case as all participants in our medium-fidelity prototyping session preferred a usable interface for the smart speaker. Smart speakers are widely used for home automation [70]. To provide users with an enhanced experience, smart speakers collect, process, and store users' data [70]. Users often find it challenging to access the privacy configuration options on the smart speaker [71, 72]. By deploying *PriviFy*, users can select the IoT device they want to configure and exercise their configuration options through physical controls. In this particular use case featuring a smart speaker, Sara uses *PriviFy* to tailor her privacy preferences as follows:

- Sara configures data retention by rotating the knob, which activates lights as she rotates that provide visual confirmation of the selected data duration. Once Sara stops at a selected time, a notification appears on the display for her to confirm her choice.

- Sara configures data usage, including targeted ads and service improvement, through the buttons. The light activates upon her enabling usage and deactivates upon her deactivating it. In each case, a notification appears on the display for Sara to confirm her choices.

- Sara configures sharing by rotating the knob. The light illuminates as she rotates it, and a notification appears to confirm her sharing choices with third parties.

## 5 Evaluation

After conducting multiple rounds of studies to ensure a comprehensive and iterative approach aligned with design considerations, we present our final prototype. The high-fidelity prototype of *PriviFy* is created using high-quality wood with lights, knobs, and button switches (Figure 4). To assess the design principles of *PriviFy*, we conducted an evaluation using multiple components of the evaluation framework [3, 73, 16], aimed at answering the following research questions:

- **RQ1-** Usability: Does *PriviFy* offer users usable privacy choices?
- **RQ2-** Findability: Does *PriviFy* make it easy for users to locate their privacy preferences?
- **RQ3-** User Engagement: Does *PriviFy* provide an engaging interface for privacy choices?
- **RQ4-** User Needs: Does *PriviFy* meet user needs in configuring their IoT privacy?

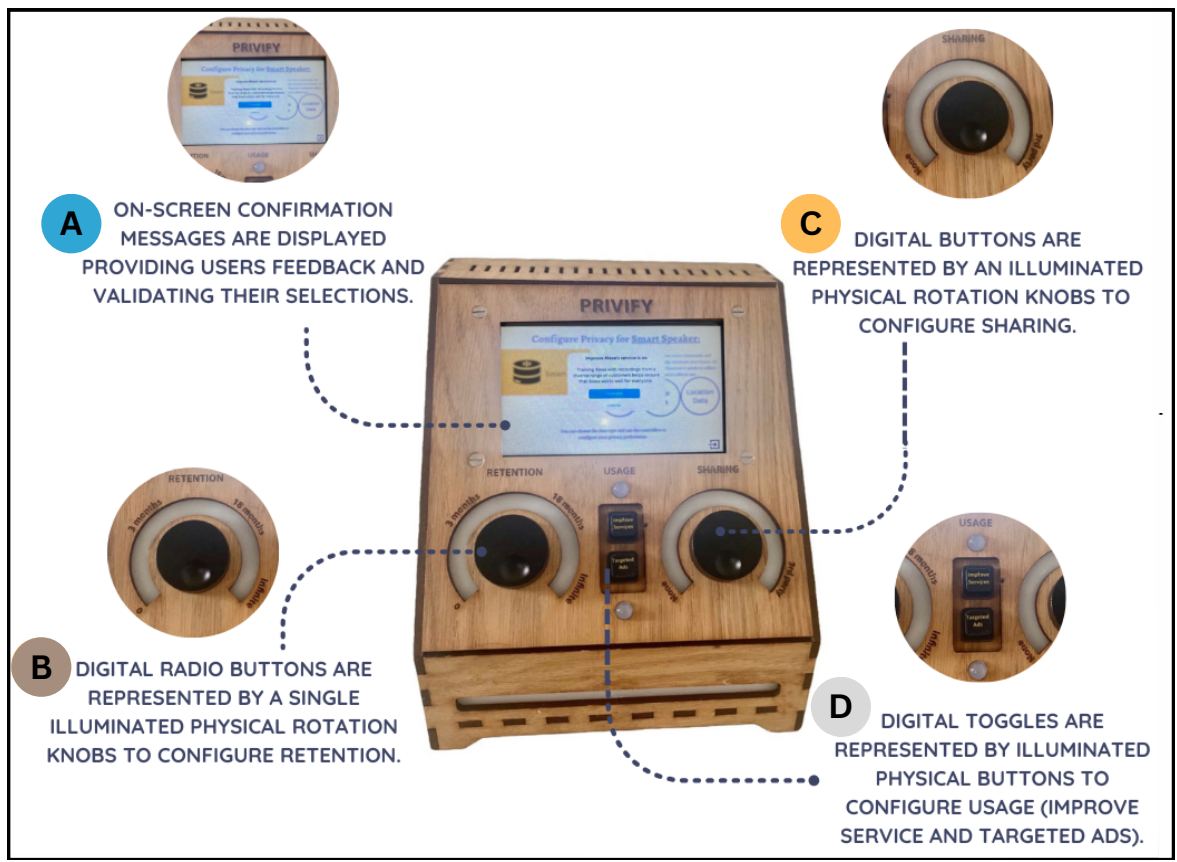

Figure 4: The high-fidelity prototype (PriviFy), illustrating privacy settings design from digital interface to tangible interface, contrasts privacy settings from multiple digital applications, Figure 1, with a tangible interface equipped with physical controls (knobs and buttons), illuminating lights, and an integrated screen. This reflects the transition towards tangible and interactive user interfaces for managing privacy preferences.

## 5.1 Participants.

We conducted a study with twenty participants recruited by advertising on the university mailing list. This number of participants is consistent with previous studies of a comparable nature [74, 24, 16]. To ensure diversity among *PriviFy* users, we extended invitations to anyone interested, specifying a preference for those owning smart speakers like Alexa in the email, albeit not a prerequisite. Our aim was to include participants with and without experience using the Alexa Application to compare their preferences and experiences. This approach allowed us to explore the potential influence of device ownership on privacy choices and participants' awareness of configuring privacy settings [71]. To avoid introducing biases, the email did not mention privacy. In a preliminary survey, participants shared demographic information. Of the participants, 11 identified as female and nine as male, with ages ranging from 26 to 40 (M = 34.5, SD = 4.89). Of our participants, 40% reported a background in information privacy, and 35% indicated that their profession was related to information privacy. Half the participants did not own an Alexa device, yet 73.8% had prior usage experience. Additionally, 50% reported never using the Alexa Application, while the remainder cited using it for work or at family/friend houses. Those who used the Alexa Application before reported using various activities, such as music control, light adjustment, speaker control, shopping, relaxation, home automation, and device setup. The semi-structured interviews took place in a university meeting room. We emailed participants before the study to schedule the time and attach the information sheet and consent form. All participants returned the consent form electronically. As an appreciation, each participant received a £30 shopping voucher after completing the study.

### 5.2 Procedure

The methodology for our evaluation was informed by similar methodologies that employed a use case-based evaluation approach [74, 75, 76]. To prepare for the study, we created a use-case scenario of a smart speaker (see Table 3 in Appendix B), prepared the tangible interface (*PriviFy*), and prepared the digital interface as the Alexa Application, which we installed on a smartphone. We configured the Alexa Application to have the necessary skills required for the use case. We conducted a within-subject study design, asking the participants to configure privacy preferences for the smart speaker use case.

We started each session by stating the study objectives, outlining the session structure, and briefly demonstrating the interfaces. Participants were then assigned two tasks: configuring privacy preferences using the tangible interface (*PriviFy*) and configuring privacy preferences using the digital interface (Alexa Application) on the smartphone, reversing the order of interface presentation. Specifically, half of the participants initiated the study with the tangible interface followed by the digital interface. In contrast, the other half began with the digital interface and then engaged with the tangible interface. This counterbalancing procedure mitigates the influence of order effects. It also strengthens the internal validity of our findings by ensuring that any observed differences are not solely attributable to the order in which we presented the interfaces. In addition, we utilized randomization methods to account for possible differences in factors such as prior knowledge of data privacy, configuration of IoT privacy settings, and familiarity with the Alexa Application [77]. This method enabled a fair distribution of participants, thus minimizing any biases related to their backgrounds, experience with IoT, and familiarity with the tools used, thereby maintaining the integrity of the study. Following the completion of each task, participants filled in the System Usability Scale (SUS) questionnaire [78], the User Experience Questionnaire (UEQ) [79], and user sentiment questions adapted from [26] and modified to fit our context (refer to Table 4 in Appendix C). In addition, participants gave qualitative feedback to evaluate the interfaces. The SUS is a standardized questionnaire comprising ten questions that employ Likert-type responses on a scale ranging from 1 to 5 to assess usability [78]. The UEQ consists of 26 bipolar items divided into six scales: Attractiveness, Perspicuity, Efficiency, Dependability, Stimulation, and Novelty to evaluate user experience [79]. The user sentiment includes seventeen questions employing a 5-point Likert scale ranging from (1: strongly disagree; 5: strongly agree).

Throughout each task, we utilized the think-aloud strategy [80], asking participants to verbalize their thoughts. We used this approach mainly to understand participants' cognitive processes and reasoning and identify any challenges they face. We thoroughly observed the participants and documented their actions and navigation strategies within the interfaces. Participants were supported in understanding task requirements and using the interface as necessary. If participants encounter difficulties completing a task, we offer hints to enhance their performance. The study duration ranged from approximately 40 to 60 minutes, with a screen recording of the digital interface interaction and an audio recording of the entire session. With the participant's permission, we took occasional pictures to document the study.

## 6 High Fidelity Prototyping Findings and Results

### 6.1 Quantitative Results

The quantitative findings demonstrate that *PriviFy* received better ratings than the digital interface. We present the outcomes of the three first research questions below.

#### 6.1.1 Usability (RQ1)

We used the SUS and the UEQ questionnaires to assess the usability between the tangible interface (*PriviFy*) and the digital interface (Alexa Application). *PriviFy* had a mean SUS score of 82.8 with a standard deviation of 9.8, while the Alexa Application had a mean of 45.7 with a standard deviation of 20.3, see Figure 5 (A). Paired t-tests indicated significant advantages for *PriviFy* at P<.05 (P = 0.000000017, t = -8.8639). Notably, *PriviFy's* SUS score is greater than the 68 thresholds for excellent usability, showing that it can provide users with usable privacy choices.

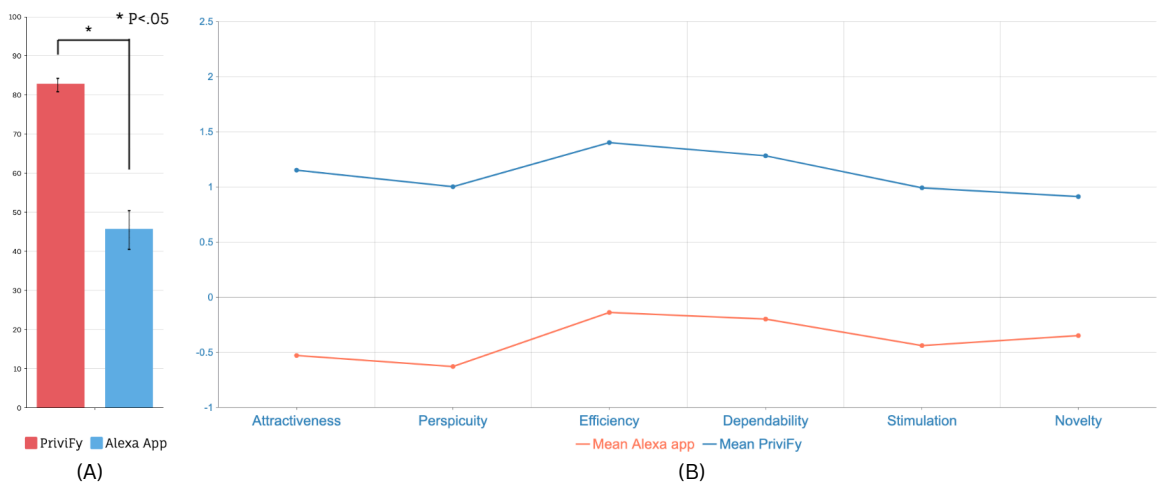

Figure 5: Results for Tangible Interface (*PriviFy*) and Digital interface (Alexa Application) of (A) SUS shown with standard error bars and (B) UEQ questionnaires.

Table 2: UEQ descriptive for Tangible interface (*PriviFy*) and Digital interface (Alexa Application) sample.

| Scale | Mean | *PriviFy* median | std | Mean | Alexa App median | std |
|---|---|---|---|---|---|---|
| Attractiveness | 1.68 | 1.92 | 0.93 | -0.53 | -0.50 | 1.15 |
| Perspicuity | 1.63 | 1.75 | 0.99 | -0.63 | -1.25 | 1.35 |
| Efficiency | 1.54 | 1.63 | 0.88 | -0.14 | -0.25 | 1.21 |
| Dependability | 1.48 | 1.50 | 0.68 | -0.20 | -0.13 | 1.15 |
| Stimulation | 1.43 | 1.75 | 1.04 | -0.44 | -0.63 | 1.10 |
| Novelty | 1.26 | 1.38 | 0.92 | -0.35 | -0.25 | 0.83 |

*PriviFy* demonstrated higher UEQ scores than the Alexa Application across all evaluated dimensions, as illustrated in Table 2 and Figure 5 (B). Paired t-tests revealed statistically significant differences in user perception for attractiveness (t = 7.545, p < 0.001), dependability (t = 6.225, p < 0.001), efficiency (t = 5.598, p < 0.001), perspicuity (t = 7.361, p < 0.001), stimulation (t = 6.338, p < 0.001), and novelty (t = 5.584, p < 0.001).

It is important to note that the evaluation of the Alexa Application was conducted with the objective of evaluating privacy choices only, i.e., the other functionalities of the app were ignored. This was also noticed by the participants. While these comments were omitted from analyses, they do explain the low ratings for the categories of the UEQ.

### 6.1.2 Findability (RQ2)

To analyze findability, we employed two metrics as follows:

**Time and path taken to find the privacy choice:** Participants took significantly longer to locate privacy choices using the Alexa Application, averaging 7.9 minutes, compared to 4.05 minutes with *PriviFy*. Our findings indicate that even participants with previous familiarity with the Alexa Application required more time to set up privacy preferences. Furthermore, several participants struggled to navigate the application, often taking a non-linear path and expressing confusion before locating privacy options. Some participants required hints to focus solely on privacy settings within the broader configuration options available in the Alexa Application. This implies that familiarity with a digital interface does not necessarily expedite navigation in such contexts. On the other hand, despite having never used *PriviFy* before, all participants demonstrated faster configuration times and more streamlined navigation, following a sequential path with minimal backtracking. Participants' performance showed no significant difference regardless of the order of interface presentation, as the time and path taken to navigate privacy choices remained consistent whether they first used the tangible or digital interface.

**Percentage of participants able to find the privacy choice:** Using *PriviFy's* interface, all participants were able to successfully find the privacy settings independently. Two participants required guidance as they moved straight to manipulating physical controls, skipping the initial step of choosing the device they intended to configure. Three participants were initially confused with the sharing knob's operation but promptly resolved their confusion upon engaging with on-screen instructions. Configuring retention posed no issues, though three participants faced challenges transitioning from activating to deactivating usage features, where they clicked on unrelated prompts. On the other hand, with the Alexa Application, all participants eventually completed the task, albeit requiring a longer duration, as discussed above. Given *PriviFy's* direct layout of privacy settings and Alexa's numerous options, we provided hints and guidance to 70% of the participants as a measure to mitigate potential biases. The guidance primarily focused on directing participants toward the location of the privacy settings. Configuring sharing posed minimal challenges to the participants. However, even with guiding the participants regarding the privacy settings location, 55% of the participants faced difficulty locating retention and usage settings. Additionally, we noticed that participants who initially began the task with *PriviFy* exhibited greater confidence in subsequent interactions with Alexa. At the same time, those starting with Alexa displayed hesitancy and sought validation for their selections.

### 6.1.3 User Engagement (RQ3)

Overall, participants showed more satisfaction engaging with *PriviFy* over the Alexa Application, refer to Figure 6. We utilized user sentiment questions to assess users' satisfaction with the interfaces through the following metrics:

**Perceptions of control after interacting with a privacy interface:** Results from our study suggest that participants generally perceive *PriviFy* more favourably than the Alexa Application in terms of providing control over their data. This preference is reflected in the higher average ratings across the five control-related questions Q2, Q13, Q14, Q15, and Q16 (Figure 6). Specifically, participants rated *PriviFy* significantly higher on Q16, with an average score of 4.25 compared to 2.9 for the Alexa Application. This difference suggests that *PriviFy* users have higher satisfaction and perceived control over their data.

**Subjective knowledge after engaging with a privacy interface:** *PriviFy* users scored higher on average across multiple metrics, indicating a better perception of informed choice (Q1), decision-making capability (Q6, Q7, and Q17), and privacy confidence (Q5). For example, *PriviFy* received a higher rating in Q6 but a lower rating in Q7 and Q17, indicating a greater perceived self-efficacy in managing privacy settings with *PriviFy* compared to Alexa's Application. The preceding suggests that physical engagement can empower users to take more active roles in data control and protection. *PriviFy's* advantage is also reflected in Q5, with an average score of 4.05, significantly higher than Alexa's average of 2.95. This highlights the importance of tangible interfaces such as *PriviFy* in building trust and confidence among users in relation to their privacy decisions, resulting in a more secure and informed data interaction experience.

**Users' comfort, trust, and investment in privacy choice interfaces:** Participants consistently expressed greater comfort and trust in their privacy choices when using *PriviFy* compared to the Alexa Application, as evidenced by their average scores on different questions. For example, when asked about the potential drawbacks of using the interfaces, *PriviFy* had an average score of 2.7 for question 19, showing a moderate level of concern, while the Alexa Application had an average score of 3.6, indicating more significant concern. This is further supported by *PriviFy's* higher average scores in Q22 and Q23, which relate to the belief that the privacy choice domain will act in the user's best interests. Participants' responses to Q22 and Q23 indicated greater trust in *PriviFy* than in the Alexa Application. Furthermore, participants' responses to Q15 indicated a higher likelihood of engagement with the privacy choice interface, with an average of 4.45, demonstrating a greater level of enthusiasm and openness to interact with *PriviFy* than the Alexa Application, which scored an average of 2.4.

## 6.2 Qualitative Results

We transcribed the interviews, conducted a thematic analysis according to prior guidelines [81], and met regularly with other authors to assess the analysis process and discuss the findings. This was mainly to address the fourth research question (RQ4) on Users' Needs and allow a more detailed understanding of the quantitative results. The qualitative analysis yields significant insights into participants' perceptions of tangible mechanisms for configuring their data privacy. Figure 7 depicts participants interacting with *PriviFy*. Below, we present the main themes derived from our qualitative analysis.

### 6.2.1 User Needs (RQ4)

As mentioned above, during the design phase, our participants identified several needs in the privacy interface. These include focusing on essential privacy options, fostering user awareness of available choices, being intuitive, reducing complexity, and expediting access to configurations. *PriviFy* demonstrated varying degrees of success in meeting users' privacy needs.

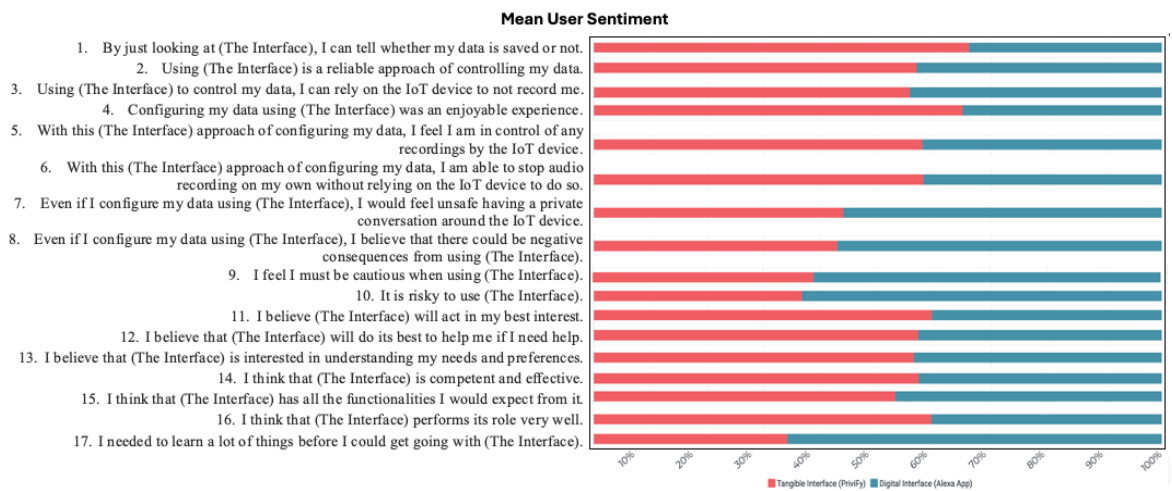

Figure 6: Results of the User Sentiment Questions, comparing the Digital Interface (Alexa Application) and the Tangible Interface *(PriviFy)*.

Participants assessed the effectiveness of the two interfaces in helping them accomplish privacy-related goals. They evaluated whether the interfaces provided the necessary tools and options to facilitate the desired privacy settings and controls. All participants expressed a preference for physical interaction when configuring privacy settings. They found *PriviFy's* physical controls more intuitive and user-friendly compared to the Alexa Application. *PriviFy* was also perceived as simple, faster, and more accessible. P6 commented, *"Even though there is information on Alexa, they are too much. I liked how PriviFy made the options clear, and the popups gave me the information I needed to understand."* Participants noted that the Alexa Application was overly complex and cluttered with numerous menus and options, leading to difficulty navigating and locating privacy settings. On the other hand, participants acknowledged *PriviFy's* effectiveness in providing the essential privacy tools and options as it encouraged them to engage with privacy settings. Some participants P2-P4, P8, P10, and P15-P17) expressed a desire for further customization on *PriviFy* beyond basic settings, mentioning a need for controls tailored to specific data types or devices. P3, for instance, highlighted the importance of granular control over data sharing. In addition, participants emphasized the potential efficacy of a dedicated tangible interface for managing privacy settings, indicating that such a feature would enhance their engagement and trust with privacy interfaces. This suggests that the tangible interface plays a crucial part in meeting users' needs.

### 6.2.2 Reliability

Participants expressed varying trust levels towards *PriviFy* and the Alexa Application concerning data privacy. Half participants showed neutral trust levels for both interfaces, while the other half preferred *PriviFy* over the Alexa Application. The participants raised ethical concerns about the transparency and consistency of data handling practices of the Alexa Application. They indicated a need for improved clarity and uniformity in the application privacy management. In particular, participants referred to the lack of consistency in prompting for user consent when turning specific privacy settings on or off. They criticized the application's approach as manipulative and potentially misleading to users. For instance, P3 expressed discomfort and quoted, *"The app only gives confirmation when I deactivate the data, but it immediately activates without confirmation when I allow data collection; I feel it is trying to steal my data."* Another participant also noted, *"I was surprised by some of the prompts and options in the interface. It made me question whether my privacy was being respected or if certain features were trying to influence my choices."*

Despite some participants preferring the Alexa Application over *PriviFy* if simplified, concerns persisted regarding its reliability in protecting data. Participants expressed more trust in tangible interfaces for privacy protection. First, participants mentioned that tangible interfaces would be mostly designed by third-party companies specialized in privacy protection, perceiving them as more reliable guardians of personal data than mainstream application developers. Second, participants emphasized that physical controls offer more security and limit the amount of captured data compared to digital interfaces. P13 commented, *"Apps are generating other data ... as a privacy-oriented person, I will try to have a physical thing which will be limited in the data and more into operations."* Third, participants see value in privacy-focused physical interfaces that offer centralized control over IoT devices and are willing to invest in such features for enhanced privacy protection. Overall, participants viewed tangible interfaces as a way to mitigate privacy concerns by limiting data collection. They emphasized the importance of trust in the interface's ability to protect user data. They also expressed a desire to control privacy settings through devices or applications explicitly designed for privacy management.

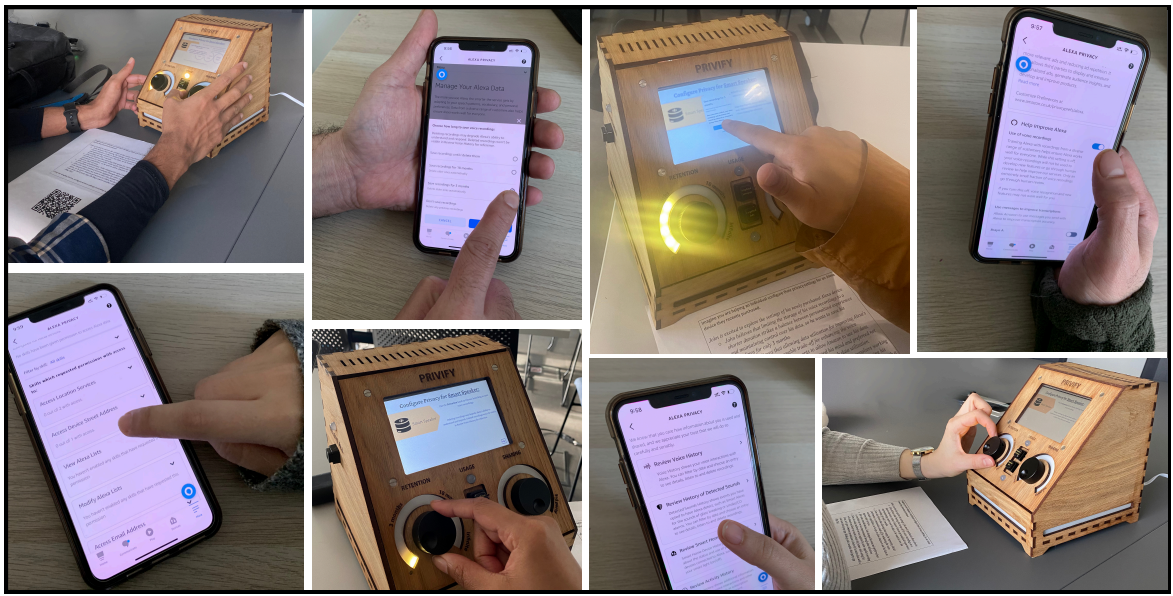

Figure 7: Multiple participants engaging in a comparison between the tangible interface (*PriviFy*) and the digital interfaces (Alexa Application). Participants engage with both modalities, including tangible knob rotation and button pressing, as well as digital radio button menus and toggles.

### 6.2.3 User Experience

Despite encountering *PriviFy* for the first time during the study, all participants displayed a notable ability to adapt to tangible interfaces. Participants favoured their experience using *PriviFy* over the Alexa Application, emphasizing *PriviFy's* intuitiveness and the ability to understand its usage without much guidance. Several participants also noted that using a tangible interface made them feel connected to the process of configuring the privacy settings. Despite the generally positive reception towards tangible interfaces, participants' feedback on the prototype presented areas for improvement, indicating an opportunity to enhance user experience through refinement. This suggests that tangible interfaces have the potential to foster interactivity and engagement when developed with careful consideration and iteration.

Enhancements proposed by participants were related to reducing the interface's size, making the buttons and knobs smaller, and improving the font size to optimize its functionality specifically for privacy settings. Moreover, suggestions were made to introduce minor adjustments, such as adding a click between each dial turn for tactile feedback, replacing the sharing knob with a switch, and enlarging the screen for improved visibility. One participant (P19) suggested streamlining the interface by only incorporating a single knob with a dynamic screen surrounding it to accommodate various devices and reduce the interface's overall size. These insights highlight valuable opportunities for refining *PriviFy's* prototype to better meet users' needs and preferences, ultimately enhancing usability and user experience in configuring privacy settings.

## 7 Discussion and Future Work

In this section, we discuss the advantages and limitations of using *PriviFy* for configuring privacy settings, as well as potential prospects for enhancing the configuration of data privacy measures.

### 7.1 Tangible Interfaces in Privacy Configuration

Findings from our analysis suggest that users prioritize privacy, value intuitive user experiences, and seek control over their data. Our results show that the tangible interface (*PriviFy*) was perceived as more effective in providing users with a sense of control over their data than the non-tangible interface (Alexa Application). This aligns with previous research, as mentioned in Sections 1 and 2, that has shown tangible interfaces to enhance user experience and engagement.

Prior research also highlights the importance of clear and transparent interface design in facilitating user understanding and decision-making regarding privacy settings. Our tangible interface prototype design received better ratings in all dimensions than the non-tangible interface. Our quantitative results reveal that *PriviFy* has better usability, findability, and user engagement than the Alexa Application. Our qualitative results also show that *PriviFy's* physical controls were better at meeting users' needs. Participants reiterated their preference for physical controls over digital applications, citing concerns about data generation and privacy infringements associated with application usage, where they viewed tangible interfaces as a way to mitigate these concerns by limiting data collection. Furthermore, the qualitative analysis emphasizes that users express more trust in *PriviFy* than in the Alexa Application, given its explicit focus on privacy management. This suggests an increasing preference for privacy-oriented features in IoT devices and the potential for *PriviFy* to enhance existing hardware with privacy-focused capabilities.

### 7.2 Enhancing Configuration Design in Privacy Interfaces

Our study investigated the efficiency of navigating privacy choices on tangible (*PriviFy*) versus digital (Alexa Application) interfaces, revealing notable differences in participant performance. Results indicated that participants took significantly longer to locate privacy options using the Alexa Application compared to *PriviFy*. Surprisingly, even participants with prior experience using the Alexa Application took longer to configure privacy settings, suggesting that familiarity with a digital platform does not necessarily expedite navigation in such contexts. Those encountering *PriviFy* for the first time demonstrated notably quicker configuration, indicating potential advantages in user intuitiveness and ease of use with tangible interfaces.

While most participants successfully located the privacy choices on both interfaces independently, varying degrees of guidance were required to navigate each interface. The analysis underscores the superior autonomy exhibited by participants when navigating *PriviFy's* tangible interface compared to the complexities encountered with the Alexa Application. Notably, the percentage of participants able to find privacy choices with minimal guidance was significantly higher when using *PriviFy* compared to the Alexa Application. These findings highlight the importance of interface design simplicity and clarity in facilitating user autonomy and efficient navigation, offering valuable insights for interface development and usability enhancement in privacy-oriented applications.

Our results show that all participants enjoyed configuring their privacy settings using *PriviFy*. This is crucial in determining the likelihood of engaging with the privacy choice interface. If users find configuring their data an enjoyable experience, they are more likely to engage with the privacy choice interface in the future. This suggests that the user experience plays a pivotal role in determining the user's willingness to interact with the interface. Positive user experiences can significantly enhance user engagement and adoption of the privacy choice interface, reinforcing the importance of a user-friendly and enjoyable interface design in fostering user trust and engagement in privacy choice domains.

### 7.3 Limitation and Future Work

*Perceptions and Objectivity.* The study's results imply that a tangible interface, exemplified by *PriviFy's*, could effectively provide users with control over their data. However, it is essential to note that these results are based on self-reported perceptions and may not fully capture the actual effectiveness of the interfaces in providing control over user data. While we tested for usability, further research using objective measures, such as behavioural data, may provide a more comprehensive understanding of user engagement and control differences between tangible and non-tangible privacy choice interfaces.

*Enhancing the Interface.* The findings also indicate that while *PriviFy* meets users' privacy needs by providing basic settings and controls, there is room for improvement in addressing diverse user objectives and enhancing the accuracy and completeness of the interface to align with user expectations and assist in goal accomplishment. These results highlight the significance of interface design in facilitating user interactions, particularly in sensitive domains like privacy settings. Further exploration into specific design elements contributing to interface efficiency could inform future interface development and user-centred design practices.

*Order of Interface Presentation.* Our study evaluation implemented a balanced approach by reversing the order of interface presentation, with half of the participants introduced to the tangible interface (*PriviFy*) first, followed by the intangible interface (Alexa Application), and vice versa for the remaining participants. Interestingly, no significant difference was observed in participant performance based on the order of interface presentation. This suggests that the order of interface presentation did not influence participants' efficiency or the time taken to configure privacy settings. This aspect highlights the robustness of the study's findings regarding the comparative efficiency of tangible versus non-tangible interfaces in navigating privacy choices.

*Potential for Specific User Groups.* Although we followed a systematic approach to design our prototype and provided a comparative evaluation between the two interfaces, our evaluation only considered one use case to evaluate the effectiveness of tangible interfaces. Our goal was to provide evidence of the ability of tangible interfaces to serve as privacy choice interfaces. To promote trust in tangible interfaces, there is a need for further research and development in the design of privacy-focused devices to meet user preferences and promote trust in IoT ecosystems. Our work holds particular potential for certain segments of society, such as older adults, neurodivergent individuals, and those with reading difficulties. These groups often prefer less screen but are drawn to visuals and physical interactions, where they may find our approach more accessible and user-friendly [82, 83]. We are aware of the potential drawbacks of designing specifically for specific groups, as it might unintentionally create barriers to other research. However, our evidence indicates that our approach benefits and is preferred by the majority of users, highlighting the significant potential of tangible interfaces in facilitating user interaction with their privacy preferences. Our work serves as an initial step towards understanding the potential of tangible interfaces in improving user control over data privacy. We aimed to ensure its usefulness, applicability, and usability. This highlights the importance of considering diverse user needs and preferences in future research and design efforts.

# 8 Conclusion

In conclusion, this paper introduces *PriviFy* (Privacy Simplify-er), a tangible interface designed to simplify the configuration of privacy settings for IoT devices. Through fidelity prototyping and iterative design with participant feedback, *PriviFy* was developed to offer users intuitive control over their data privacy preferences. Our evaluation demonstrated that *PriviFy* significantly enhances privacy control among users, enabling them to configure their privacy settings in a user-friendly way. The findings emphasize the critical role of user-friendly interfaces in addressing the challenges associated with managing privacy preferences for IoT devices. By providing clear and accessible controls, *PriviFy* empowers individuals to regain control over their data privacy, thereby mitigating concerns related to transparency and complexity in existing privacy controls. In summary, *PriviFy* represents a promising approach to addressing the complexities of privacy settings configuration for IoT devices. Continuing research, design, and regulation efforts are essential to promote user autonomy and enhance privacy protections in an increasingly connected world.

# A   Fidelity prototyping

## A.1   The three low-fidelity prototypes developed by the research team are as follows:

- A cuboid shape with 5 sides. The top side has a touchpad, where the user could swipe right and left to show the data collected by a specific IoT device. The other four sides has a mixture of buttons and sliders to control the configuration of collection, usage, sharing, and retention.

- A table layout with two sides. One side has a touchpad, where the user could swipe up and down to show the data collected by a specific IoT device. The other side has 4 sections with a slider, where the user could slide to control their privacy preferences. The 4 sections are the configuration of collection, usage, sharing, and retention.

- A table layout with two sides. One side has a card reader, where the user could place a card for each IoT device, configure their data on the other side, and then save their privacy preferences. The other side has the same sections as the second prototype.

## A.2   The two medium-fidelity prototypes developed by the research team are as follows:

- A design resembling an ATM, featuring two sides. The primary interface features a touchpad that allows the user to select the desired IoT device for configuration. Additionally, it has four configuration buttons, with two located on each side of the touchpad. Below the touch pad, there are three control buttons for configuring privacy preferences. The other side includes a "delete all" button that enables the user to erase all of their data instantly.

- A control panel design with 2 sides. The primary interface features a touchpad that allows the user to navigate through IoT devices and display the collected data by swiping right or left. Additionally, it has three configuration options, two of which are controlled by a rotating knob and one by two clickable buttons. The other side includes a "delete all" button, which is similar to the aforementioned prototype.

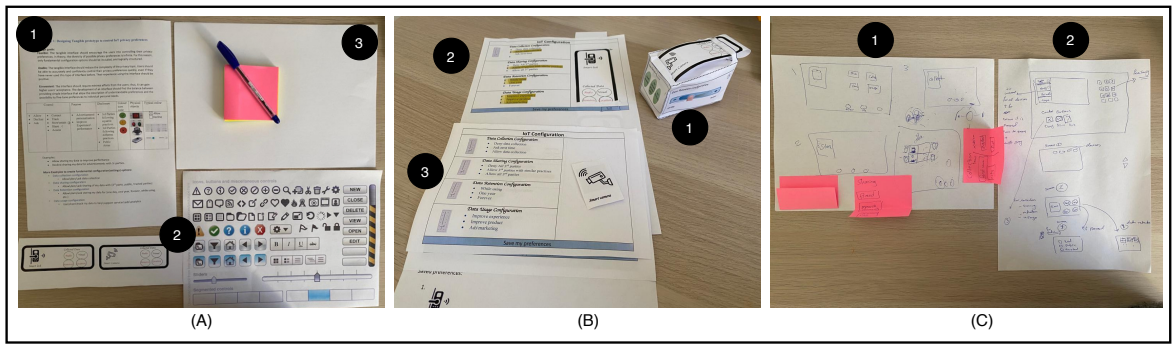

Figure 8: Study Materials and Prototypes in Design Iterations:(A) Study materials distributed to participants for research engagement and data collection: (1) Guidelines page, (2) Notes and pens, and (3) Prototyping sheet. (B) Three low fidelity prototypes developed by the research team. (C) Two low fidelity prototypes developed by the participants, contributing to the iterative design process.

Table 3: Smart Speaker Use Case presented to the participants so they can configure the Privacy preferences using the two interfaces.

| Smart Speaker Use Case |
| --- |
| Imagine you are helping an individual configure their privacy settings for an Alexa device they recently purchased. |
| John is excited to explore the settings of his newly purchased Alexa device. o John believes that limiting the storage of his voice recordings to a shorter duration strikes a balance between personalized experiences and maintaining control over his data, so he wants to save his voice recordings for only 3 months. o John also believes that allowing data utilization for improving Alexa's performance is a reasonable trade-off for enhancing the voice assistant's capabilities, so he opts to allow Amazon to use his data. o After some consideration, John changed his mind and preferred not to contribute to service improvements through data utilization. o John also values tailored services supplied by other suppliers working with Amazon and does not want to miss this feature while using his Alexa, so he gave Amazon permission to share his location data with a ride-share third-party app (BeMup) and a weather third-party app (Big Sky). |
| Using Alexa App/*PriviFy*, configure privacy settings according to the scenario. |

## B   use case

This section presents the use case used in the evaluation of the study.

## C   IoT devices evaluation

This section presents some of the evaluation materials that were used in the study.

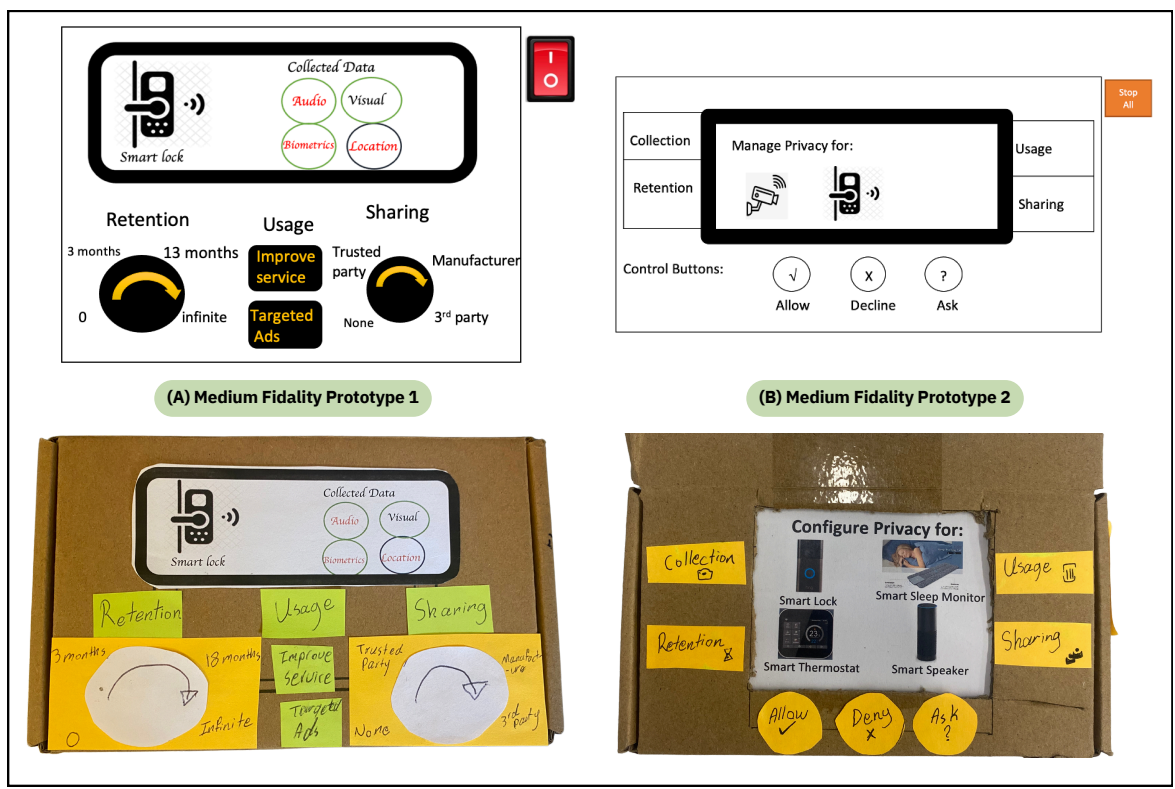

Figure 9: Two Medium Fidelity Prototypes Developed by the Research Team, Presented in Digital and Physical Formats. Each prototype is depicted in two images: one showcasing the digital representation, while the other presents the physical format made from cardboard and stickers for user interaction. This medium-fidelity approach provides tangible representations for user engagement and feedback collection.

Table 4: User Sentiment Questions given to the Participants: the Term (The Interface) is modified to *PriviFy* or the Alexa Application.

| User Sentiment Questions |
| --- |
| By just looking at (The Interface), I can tell whether my data is saved or not. |
| Using (The Interface) to control my data, I can rely on the IoT device to not record me. |
| Configuring my data using (The Interface) was an enjoyable experience. |
| With this (The Interface) approach of configuring my data, I feel I am in control of any recordings by the IoT device. |
| With (The Interface) approach of configuring my data, I am able to stop audio recording on my own without relying on the IoT device to do so. |
| Even if I configure my data using (The Interface), I would feel unsafe having a private conversation around the IoT device. |
| Even if I configure my data using (The Interface), I believe that there could be negative consequences from using (The Interface). |
| I feel I must be cautious when using (The Interface). |
| It is risky to use (The Interface). |
| I believe (The Interface) will act in my best interest. |
| I believe that (The Interface) will do its best to help me if I need help. |
| I believe that (The Interface) is interested in understanding my needs and preferences. |
| I think that (The Interface) is competent and effective. |
| I think that (The Interface) has all the functionalities I would expect from it. |
| I think that (The Interface) performs its role very well. |
| I needed to learn a lot of things before I could get going with (The Interface). |

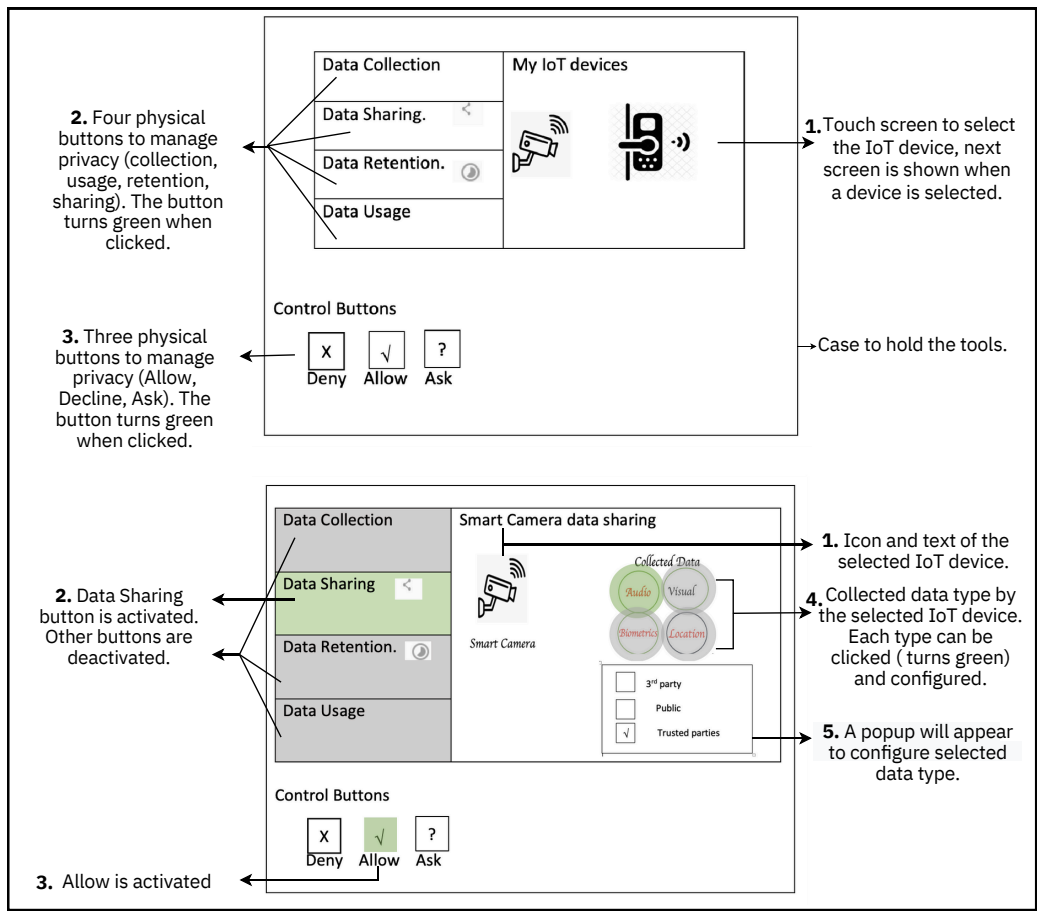

**(A) Group 1 Low Fidality Prototype**

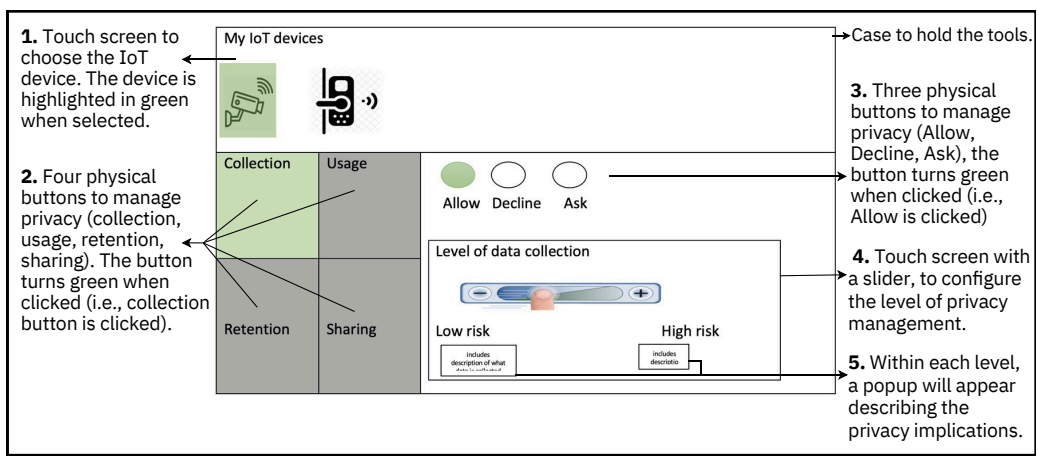

**(B) Group 2 Low Fidality Prototype**

Figure 10: Computerized Renderings of the Two Prototypes Developed from Participant sketches in Figure 8 (C). The prototypes were initially sketched by participants on paper and subsequently converted into digital format for enhanced clarity.

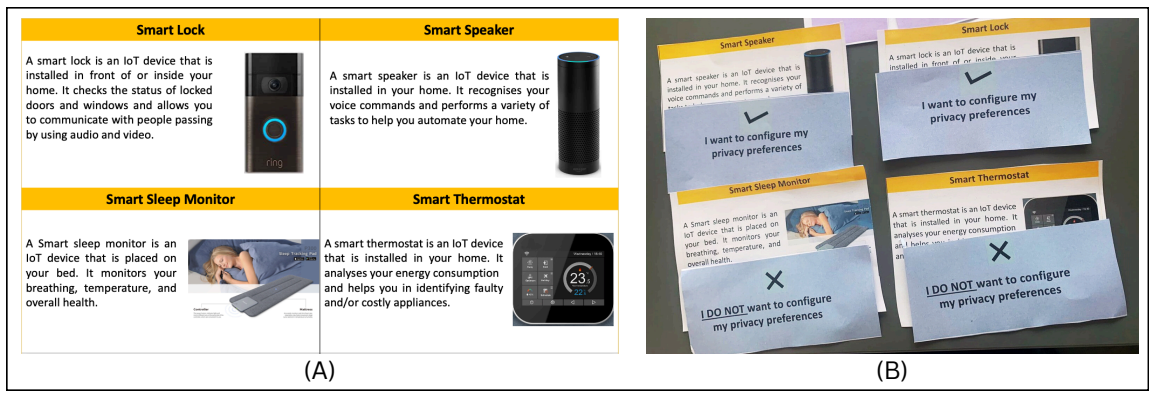

Figure 11: Use cases used in Medium Fidelity Prototyping. (A) Four different use cases of IoT devices are outlined to the participants. (B) Participants actively selecting their preferred use case for configuring privacy settings, reflecting their involvement in the customization process.

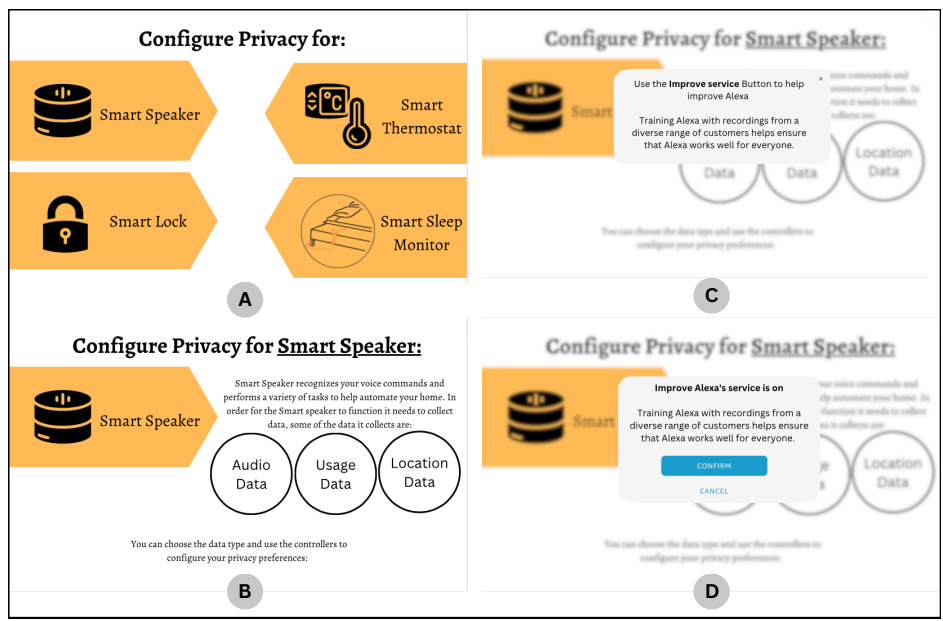

Figure 12: Interface Screenshots of the High Fidelity Prototype for IoT Device Configuration. (A) Depicts the home screen presenting users with device selection options. (B) Illustrates configuration choices for a smart speaker use case. (C) Displays a notification helping users in locating physical controls for data management. (D) Shows confirmation messages providing feedback to users. These screenshots offer insight into the user interface design and functionality of the prototype.